\documentclass[sigconf]{acmart}

\AtBeginDocument{%
  }

\copyrightyear{2026}
\acmYear{2026}
\setcopyright{cc}
\setcctype{by}
\acmConference[CI '26]{Proceedings of the ACM Collective Intelligence Conference}{September 27--30, 2026}{Alexandria, VA, USA}
\acmBooktitle{Proceedings of the ACM Collective Intelligence Conference (CI '26), September 27--30, 2026, Alexandria, VA, USA}
\acmDOI{10.1145/3834581.3838629}
\acmISBN{979-8-4007-2895-2/2026/09}

\usepackage{array}
\usepackage{multirow}

\begin{document}

\newcommand{\frameworkname}{\texttt{Topic}-\texttt{Discourse}-\texttt{Reasoning Framework}}
\newcommand{\frameworkshortname}{\texttt{TDR Framework}}
\newcommand{\edit}[1]{\textcolor{blue}{#1}}
\newcommand{\sssec}[1]{\textbf{#1}}
\newcommand{\tabitem}{~~\llap{\textbullet}~~}

\newcommand{\topic}{Topic}
\newcommand{\topics}{Topics}
\newcommand{\discourse}{Discourse Role}
\newcommand{\discourses}{Discourse Roles}
\newcommand{\reasoning}{Reasoning Type}
\newcommand{\reasonings}{Reasoning Types}



\title[Disentangling Threads]{Disentangling Threads: Exploring the Potential of LLM-Supported Discussion Forum Analysis for Community Insight}

\author{Tony Li}
\email{toli@ucsd.edu}
\orcid{0000-0001-7552-6689}
\affiliation{%
  \institution{University of California, San Diego}
  \country{USA}
}

\author{Zhiqing Wang}
\email{zhw131@ucsd.edu}
\orcid{0009-0008-6642-9092}
\affiliation{%
  \institution{University of California, San Diego}
  \country{USA}
}

\author{Thanh-Nha Tran}
\email{tntran@ucsd.edu}
\orcid{0009-0005-4929-2109}
\affiliation{%
  \institution{University of California, San Diego}
  \country{USA}
}

\author{Yu-Chun Grace Yen}
\email{yyen@nycu.edu.tw}
\orcid{0000-0002-5442-6934}
\affiliation{%
  \institution{National Yang-Ming Chiao-Tung University}
  \country{Taiwan}
}

\author{Steven P. Dow}
\email{spdow@ucsd.edu}
\orcid{0000-0002-1354-9866}
\affiliation{%
  \institution{University of California, San Diego}
  \country{USA}
}

\renewcommand{\shortauthors}{Li et al.}

\begin{abstract}
Online discussion forums enable people from diverse backgrounds to share ideas, feedback, and perspectives. These organic discussions can help researchers understand communities’ collective viewpoints, but insights are often difficult to uncover given their freeform reply structure. Large language models (LLMs) support qualitative text analysis but can misalign with researchers' analytical intent and miss key insights. To inform design considerations for forum sensemaking tools, we manually analyzed a forum discussion, synthesized an exploratory analysis framework from relevant literature, built a design probe, and interviewed 21 researchers to uncover perceived opportunities and barriers with LLM representations of collective discussions. We provide recommendations for community sensemaking tools to support flexible analytical goals grounded in raw user data and enable follow-up research processes, while balancing anonymous free expression with the desire for contextual information on commenters.
\end{abstract}




\maketitle

\section{INTRODUCTION}
Online discussion forums (e.g., Reddit, HackerNews) empower
researchers\footnote{We use ``researcher'' to refer to anyone interested in understanding collective viewpoints, e.g., community designers, policymakers, social system researchers.}
to examine how communities collectively make sense of complex issues.
The evolution of topics, viewpoints, and interaction dynamics within threaded conversations over time offers a unique lens for collective sensemaking.
Online forums can provide more insight on the diversity of personal opinions within a community than offline methods such as interviews \cite{jamison2018online}.
Other methods like surveys can also capture a wide range of community opinions, but they can suffer from low-quality responses and nonresponse \cite{porter2004survey, xiao2020survey}, and
they don't explicitly allow people to interact with each other or implement solutions \cite{mahyar2018communitycrit}.
In contrast, forums can offer researchers an accessible snapshot of community discourse as a complement to primary research methods like interviews and surveys.
However, forum analysis is a manual and resource-intensive process that can be challenging for rich discussions \cite{jiang2021serendipity, robinson2014subsampling}.

Large language models (LLMs) promise efficient support for text analysis \cite{byun2023llm, carius2024llm, lam2024lloom} and qualitative insights \cite{dovetail_tools:online, atlasti:online, gebreegziabher2023patat, gao2024collabcoder}, but they are known to miss semantic nuance \cite{carius2024llm, romberg2024review} and produce biased outputs \cite{liang2021bias, nadeem2021bias}.
Furthermore, the tangled nature of online discussions often involve related ideas scattered across sub-threads, conversations that drift off-topic, and sprawling hierarchical structures that researchers must navigate without support \cite{macneil2021designspace, latkovikj2020forums}.
Forum comments themselves are also freeform, potentially irrelevant, and meaningful only in the context of other comments and community norms \cite{macneil2021designspace, latkovikj2020forums}.
As AI adoption increases, a growing body of research explores how LLMs assist (and hinder) researcher understanding \cite{schroeder2025llm, liao2024llm, jiang2021serendipity}. 
In this work, we seek to gain empirical insights into how researchers perceive the the potential of LLM-supported tools for analyzing forum discussions for collective intelligence research. We explore the following two research questions:

\begin{itemize}    
    \item [\textbf{RQ1}:] What opportunities do researchers see for LLM-assisted analysis of discussion forums?
    \item [\textbf{RQ2}:] What barriers do researchers perceive around using LLMs for forum-based community research?
\end{itemize}

To address these questions, we (1) manually analyzed a real forum thread and synthesized a novel \frameworkname\ from prior work on topics \cite{gebreegziabher2023patat, gao2024collabcoder}, discourse roles \cite{zhang2017discourse, zakharov2021discourse, kolhatkar2020constructive}, and reasoning types \cite{xia2022persua, irani2024argusense, kolhatkar2020constructive} to develop a design probe, (2) conducted a preliminary LLM feasibility analysis for automatically extracting these dimensions, and (3) interviewed 21 designers experienced in qualitative analysis to elicit their viewpoints on LLM-assisted forum analysis.

Participants expressed optimism about using forums both for exploratory research and for targeted research questions.
They strongly desired interacting with grounded data for their varied needs but also to inform their follow-up and complementary research.
On the other hand, participants raised concerns about comment quality, commenter credibility, and LLM semantic accuracy.
We also observed a key tension: researchers mentioned the importance of contextualizing analysis with information about commenters, but they recognized that adding identity markers could take away from the authenticity of an open and anonymous discussion.
We discuss design recommendations for future LLM-assisted tools that
ground analysis with real data, balance commenter context with open expression, and enable follow-up primary research both within and informed by forums.
\section{RELATED WORK}
We surveyed the literature
on online discussions' analytical potential and the tools for LLM-assisted text analysis to understand the research landscape and limitations.

\subsection{The Potential Research Value of Online Discussions}
Online discussion forums offer a low-cost, open-ended way for people to discuss and solve problems \cite{latkovikj2020forums, macneil2021designspace, liu2023coargue, reynante2021civic}. They also have the potential to provide researchers with insights into unfiltered perspectives from diverse participants \cite{smedley2021practical}, as a secondary research method that complements approaches like interviews and surveys \cite{jamison2018online, lee2017understanding}.
However, making sense of large bodies of forum text is time-consuming \cite{jiang2021serendipity, robinson2014subsampling}, with information distributed across threads and duplicate ideas \cite{macneil2021designspace}.

Commercial social-listening tools track forum trends for market insights and social media branding with aggregate metrics such as sentiment measurement and topic keyword prevalence \cite{brandwatch:online, sproutsocial:online}, but they lack functionality for deeper collective intelligence research such as understanding the range of perspectives or designing solutions for emergent community needs.
Prior work explored visualizations for summarizing the topic space in a discussion based on crowdsourced interpretations, including
for civic insight \cite{kriplean2012considerit, faridani2010opinionspace, chen2006viz, jasim2021communitypulse}, users' opinions on social media \cite{song2023needs, andre2014synthesis, zhang2017wikum}, and collaborative tagging to make sense of unstructured chats and comments \cite{zhang2018tilda, willet2011commentspace, yen2020decipher}.
However, crowdsourcing methods can require significant costs to recruit people to summarize a sprawling discussion, and given recent advances in AI, the focus has turned to using LLMs and LLM-supported tools for analysis.



\subsection{The Rise of LLM-Assisted Qualitative Text Analysis}
LLMs have been increasingly utilized for qualitative understanding, including in thematic analysis, codebook creation, and collaborative qualitative coding \cite{byun2023llm, carius2024llm, lam2024lloom, gebreegziabher2023patat, gao2024collabcoder, overney2024sensemate, depaoli2024inductive, jiang2021serendipity}, and appears especially suited for deductive coding that leverages pre-determined codebooks \cite{tai2024llm, xiao2023deductive, rao2024quallm, marathe2018qual, chew2023deductive}.
Commercial tools such as Dovetail and ATLAS.ti also offer LLM features to propose inductive qualitative codes and summarize data \cite{dovetail_tools:online, atlasti:online, shen2023llmresearch}.
However, LLMs often struggle with semantic context and expressive language \cite{gebreegziabher2023patat, carius2024llm, romberg2024review} and can produce biased outputs \cite{liang2021bias, nadeem2021bias}, making their application to freeform, subjective, and tangled forum data uncertain.
Researchers have also criticized the use of LLMs for qualitative analysis, citing the importance of having a human interpretative lens \cite{jowsey2025qual}. 

Given the limitations of current approaches and the potential innovation space for LLM-based qualitative sensemaking,
it becomes timely to empirically understand researchers' analytical needs and perceived limitations of existing LLM technologies for community forum analysis. Ultimately, the goal is to design analytical tools that empower rather than replace human researchers, specifically to help them disentangle the complexity of forum discussions.

\section{DEVELOPING OUR GROUNDED PROBE}
\begin{figure*}[htbp!]
  \includegraphics[width=\textwidth]{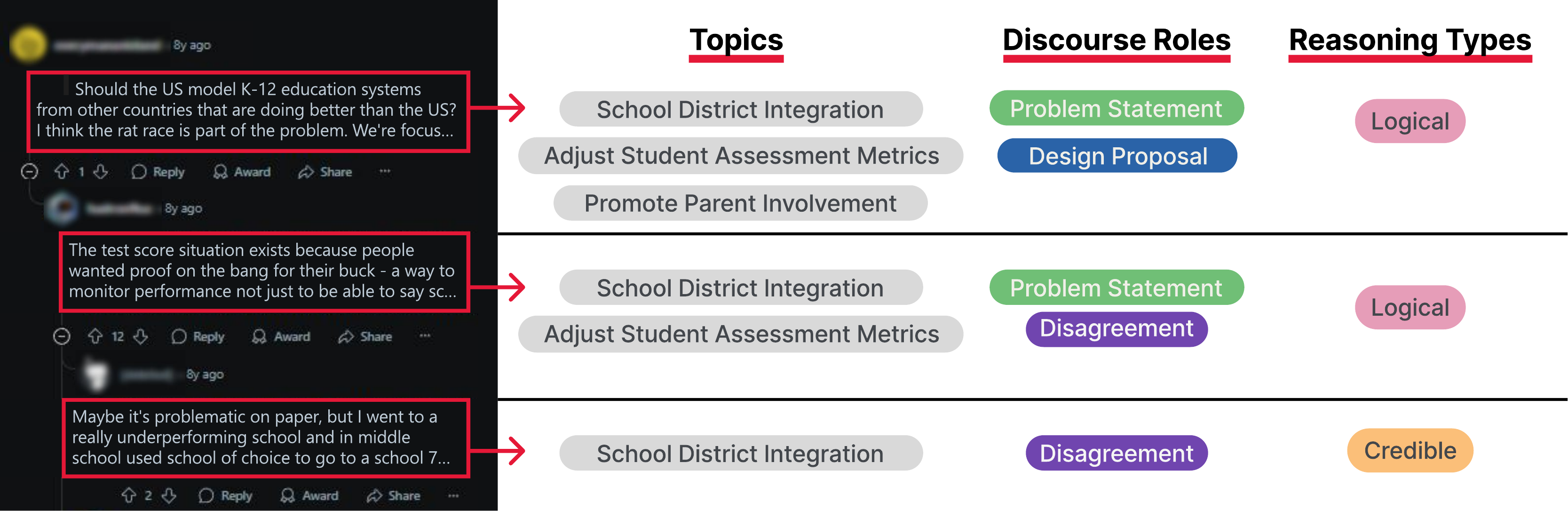}
  \caption{\frameworkname\ labels for sample comments in a subreddit thread for policy design \cite{redditthread:online}. Comments have been truncated, and usernames and profile pictures in the Reddit screenshot have been blurred to protect user privacy.}
  \Description{A screenshot of comments from a subreddit on the left side, annotated with labels of topics, discourse roles, and reasoning types for each comment on the right side.}
  \label{fig:example}
\end{figure*}

Inspired by design probes in HCI research \cite{gaver1999probe, dow2016probe}, we designed a probe demonstrating the potential analytical capabilities of LLM-enabled forum analysis to ground reflections on our research questions.
We manually analyzed an actual forum discussion, synthesized a \frameworkname\ from related work, explored if an LLM workflow could feasibly derive the framework dimensions, and built a prototype interface as the design probe.


\subsection{Analytical Dimensions of Online Discussions}\label{sec:framework}
Prior LLM work has focused on surfacing dominant themes in texts such as academic papers and news articles, rather than the interactive and evolving discourse that is unique to online discussion forums.
In order to explore forums' potential richness for community insight, we expanded our
sensemaking lens beyond simple topic analysis
to explore analytical dimensions that could help unpack the collective knowledge expressed in forums.
Prior discussion analysis work categorizes comments by \topic\ (the ideas discussed) \cite{gebreegziabher2023patat, gao2024collabcoder, lam2024lloom}, by communicative \discourse\ \cite{zhang2017discourse, kolhatkar2020constructive}, and by persuasive \reasoning---Logos (logic), Ethos (credibility/experience), and Pathos (emotion) \cite{xia2022persua}.
We adapted Zhang et al.'s discourse roles \cite{zhang2017discourse} for design and research contexts as Problem Statement, Design Proposal, and Agreement/Disagreement,
since these dimensions were echoed in other systems \cite{khartabil2021argviz, liu2018consensus, jasim2021communitypulse, hoque2016multiconvis}, as opposed to other discourse roles such as Humor or Elaboration.
These dimensions and their intersections
consolidate into our novel analytical schema, \frameworkname. Our research objective here was not to validate the \frameworkshortname\ but to use it as a probe to understand if and how researchers could make sense of forum dimensions for community insights.

\subsection{Prototype Design}
To provide visual structure for the design probe interviews with researchers, we created a prototype analysis dashboard UI that leverages the \frameworkshortname\ for the sample forum discussion.
This was meant to ground participants in a research task to enable them to reflect on contextual and technical details they might not be able to recall without a grounded probe \cite{gaver1999probe}.

\subsubsection{Data 
Source}\label{ssec:reddit_dataset}
We sourced data from a public Reddit thread, ``How can the US improve its K-12 education system?'' \cite{redditthread:online}---an open-ended design question with roughly 320 comments spanning multiple perspectives over several years. We requested research access directly through Reddit and followed their developer and data policies for collection and analysis. The study protocol with this data was also reviewed and determined exempt by our Institutional Review Board.

Two authors independently performed inductive open coding for \topics, discussed and merged a codebook, then re-coded until agreement stabilized through iterative, reflexive discussion and note-taking.
See \autoref{append:topic_codebook} for this final codebook of \topics.
The same authors also independently applied deductive labels for \discourse\ and \reasoning\ following our predetermined breakdowns (Section~\ref{sec:framework}), resolving discrepancies through iterative discussion and note-taking.
The mutual intersection of the authors' tags comprised the final dataset labels.

For the probe, we populated the prototype with these manually labeled data but asked participants to imagine the information was LLM-analyzed, to probe perceptions of a scaled-up LLM implementation.
See \autoref{fig:example} for example labeled comments.


\subsubsection{Prototype Interface}
\begin{figure*}[h]
    \centering
    \includegraphics[width=1.0\linewidth]{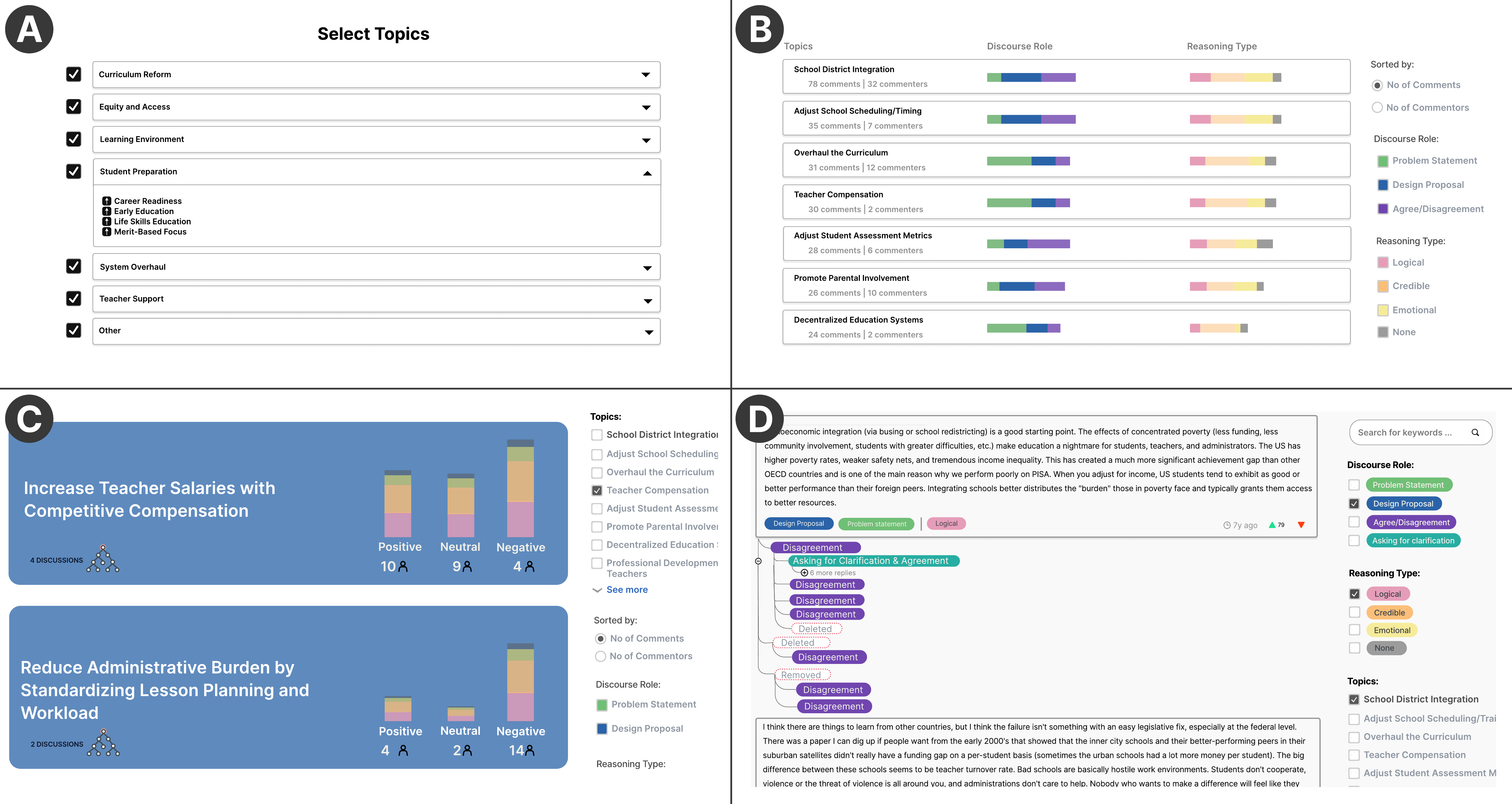}
    \caption{Pages in our probe: (A) Topic Selection; (B) Topic Overview; (C) Subtopic View; (D) Comment View}
    \label{fig:probe_screens}
    \Description{Four screenshots: a list interface listing out expandable topics, banners of topic names with colored stacked bar charts of comment breakdowns, colored banners of subtopics with stacked bar charts of comment breakdowns, and bubbles of comments in a reply tree structure labeled with colored comment type badges.}
\end{figure*}

Referencing discussion sensemaking literature \cite{song2023needs} and existing discussion visualizations \cite{hoque2016multiconvis, hoque2015hil, wu2014viz, hu2013viz, jasim2021communitypulse, liu2018consensus},
we designed the probe interface as a dashboard presenting the labeled discussion through four complementary views: \textit{Topic Selection} (checklist of AI-generated topics for human oversight and editing), \textit{Topic Overview} (aggregate comment counts and \discourse/\reasoning\ distributions per topic), \textit{Subtopic View} (reply-tree breakdowns within a topic), and \textit{Comment View} (hierarchical replies with color-coded TDR tags collapsible by dimension, with an optional AI-summary toggle to view LLM-generated label justifications) (see \autoref{fig:probe_screens}).
We kept the prototype static (non-interactive) so participants could focus on the analytical affordances rather than system usability. Since this was an exploratory probe of researchers' mindsets, we opted for more controlled elicitation over technical fidelity.

\subsection{Preliminary LLM Feasibility}
We implemented an LLM workflow adapted from LLooM \cite{lam2024lloom} to inductively generate topics and classify comments across all three TDR dimensions.
A separate technical evaluation found that bottom-up LLM \topic\ generation covered 89\% of our manual topics, and the LLM classification of \frameworkshortname\ labels outperformed BERT, RoBERTa, and logistic-regression baselines on macro-F1 scores across all dimensions, though there was still room for improvement compared to the manual labels themselves (full workflow, prompts, and evaluation tables are available in supplemental material). While the feasibility analysis only covered a single forum discussion, it provided preliminary indication that LLMs could interpret these dimensions in a way that aligns with human interpretation and can scale more accurately than existing natural language models.

\section{METHOD}\label{sec:design_probe}
We conducted interviews with 21 researchers using our probe to elicit concrete reflections on the opportunities and barriers of LLM-assisted forum analysis for community research. 


\subsection{Participants}
We recruited participants through our institution's communications channels (namely emails and Slack announcements) and internal subject pool recruitment system,
supplementing our recruitment with several personal network reach-outs to include more experienced qualitative analysts.
The screened pool of 21 participants all had experience with qualitative data analysis projects.
18 participants had experience posting in online discussion threads.
18 participants identified as female, and the remaining identified as male.
14 participants were between the ages of 18 and 24, 6 were 25-34 years old, and the remaining was 35-44 years old.
Participants were compensated at a rate of \$20 USD/hr or course credit if they were a current student through the internal recruitment system, compliant with our IRB.

\subsection{Design Probe Procedure}
Conducted as a 1-hour semi-structured online interview (see \autoref{append:interview_guide} for the interview guide), the study protocol comprised four stages:

\sssec{Pre-Study Questionnaire and Warm-Up (5 min).} Participants completed a pre-study questionnaire and consent form that briefed the goals of the study and assessed their prior experience and familiarity with design practices and online discussion forums. At the beginning of the interview, participants were asked to confirm their consent before answering warm-up questions detailing their responses on this questionnaire. This step provided background information to contextualize their interaction with discussion forums and qualitative data analysis.

\sssec{Baseline Task (15 min).} Participants first completed a baseline task, which involved freely exploring the Reddit discussion thread related to K-12 education improvement \cite{redditthread:online} while thinking aloud on their thread insights.

\sssec{Probe Tasks (30 min).} Following the baseline task, participants engaged in structured think-aloud probe tasks designed to assess specific visualization features of our prototype, focusing on interactions with each \frameworkshortname\ dimension, as well as the subtopic view and AI-summarized views of the prototype. For each feature, participants were asked to discuss their expectations, relevance to their decision-making process, and potential suggestions for additional functionality.

\sssec{Post-Study Interview (10 min).} After completing the probe tasks, participants answered questions on three main topics: (1) insights gained during the tasks that could inform policy design decisions, (2) feedback on specific design prototype features, and (3) additional information needs or functionalities that could further enhance decision-making.



\subsection{Interview Analysis}
Two authors analyzed the interview transcripts and recordings via inductive thematic analysis \cite{braun2006thematicanalysis}.
First, they reviewed the same subset of 3 interviews and independently generated codes, meeting to review the codes and discuss the logic for a codebook and how to add new codes.
The authors then independently coded two subsequent interviews at a time, reconvening to compare codes and discuss until reaching code saturation, which was achieved after four iterations. They then divided the rest of the interview data for analysis, meeting up to discuss unclear edge cases that arose, mutually agreeing on changes to the codebook and re-coding until the codebook was stabilized.

With the codebook of around 350 codes organized into 12 code groups, the two authors inductively grouped codes into clusters of concepts. The authors then independently derived themes from those concepts, reflexively discussing and comparing ideas before reaching consensus on the final set of themes.

\section{RESULTS}
Our thematic analysis revealed insights around participants' desires and hesitations on LLM functionalities\footnote{Participants mentioned the general term ``AI'' with respect to modern natural language analysis and generation, so we interpreted these comments as referring to LLMs.}.
In general, participants applied forum comments to inform both exploratory and targeted community insights, but they desired more analytical control, demographic context for commenters, and source-grounded verification for LLM outputs.

\subsection{RQ1: What opportunities do researchers see for LLM-assisted discussion forum analysis?}
Participants valued high-quality comments from diverse users in discussion forums.
Through the probe, they articulated how they would find those desired comments and reflected on additional analytical functionalities,
across a range of
early-stage exploratory research goals
and later-stage validation or targeted searches,
as well as follow-up researcher actions.

\subsubsection{Understanding diverse perspectives to guide exploratory research.}
Nearly all participants highlighted the value of hearing different commenters' inputs on forums, especially personal anecdotes and viewpoints that evolve through discussion.
They agreed that ``it's nice hearing actual opinions'' (P17) and ``more potentially honest or candid feedback'' (P21)
on forums. Participants also valued how forum platforms allow for more diverse people to converse, mentioning that
``you get such a variety of people [in forums], but it is a solid congregation of people [...] there really aren't other places that you can find such a wide variety of information'' (P20),
especially ``to build like a world rounded-ness on how the public is feeling'' (P10).

For \topics, participants wanted to aggregate similar insights across different comments.
For instance, when participants analyzed the raw discussion forum in the baseline task, they mentioned wanting ``some sort of intelligent system [...] to say, okay it seems like these three comments are similar---they're talking about the same thing'' (P1) and even an explicit ``tagging system [...] a bird's eye view of what topic they're talking about'' (P6).
After interacting with the probe, participants appreciated the ability to see the full breadth of all topics mentioned in a discussion. They mentioned that ``topics kind of give me some sort of guide for my exploration [... so I] don't miss out on any of the insights that pertain to these topics'' (P1), and that they would appreciate being able to see ``data that I may not necessarily have already some hypotheses, or some idea of what to be looking at'' (P21).
A couple participants even mentioned wanting to be able to see insights aggregated beyond a single discussion, such as ``data from different posts together'' (P3) and ``other running threads that are related [...] not limiting myself to just one single Reddit thread'' (P8).
In addition, with the \discourse, participants wanted to quickly understand aggregate sentiment for both sides of arguments around topics, to ``see how many people agree with the topic'' (P18) and ``the most receptive posts'' (P1).

\subsubsection{Developing a deeper understanding of design scenarios and disagreements to inform targeted inquiries.}
Participants also valued research insights from the forum beyond exploratory understanding.
For example, participants wanted to ideate and test out ideas in forums, mentioning they could pursue a ``scientific process'' to test their ``hypothesis of what the problem is'' and their ``design proposal that I have''
(P10). Furthermore, a couple participants specifically mentioned how the framework could be embedded in their existing qualitative analytical process to revisit data, such as for finding representative quotes:
``When I do the analysis, I remember, oh I heard this, or I read this, but then I miss where it is, so I want to look it up and jump to it'' (P4), and
``in industry there's so many shortcuts that a lot of times things aren't rigorously tagged like that. The fact that these [framework labels] kind of like supplement maybe not having been done correctly the first time is really helpful. Really, really helpful'' (P20).

Participants referenced the \frameworkshortname\ in the probe to help frame their mental models of what they sought in discussions, with participants mentioning
``it's kind of categorized in a certain way, I kind of know what I'm looking at already'' (P12), even if they had not thought of these breakdowns in design terms before: ``it's interesting using this [framework's] phrasing because I think it actually does align well, but I hadn't thought of these in terms of, like, design wording of Proposal, Problem Statements... but I guess that is what it is'' (P20).
All participants felt the conceptual framework demonstrated in the design probe would help them find desired information for design decision-making, but specific goals varied---from guiding exploration of the topic space, to surfacing debate conversations around a particular issue, to screening out target comments based on their conversational and contextual roles.
\topics\ helped participants browse specific areas of interest, \discourse\ helped identify potential solutions and feedback for those topics, and \reasoning\ contextualized comments for interpretation.
See \autoref{tab:dim_goals} for participants' most commonly-mentioned use cases for each \frameworkshortname\ dimension.

Participants also noted the distinct value in combining different aspects of the framework, both reinforcing their prior mental models of what information is important and helping them discover useful slices of the data. In particular, participants noted cumulative value, such as with a sequential analytical process through ``the detailed, decomposed topics, and I want to see how many positive and negative and net neutral opinions on that. And then I can focus more on the reasons or more on the problem statements'' (P6).
In addition, combining framework dimensions brought more value than any alone, with participants mentioning ``it's important to kind of look more at the logical and credible side by side, because if someone is logical but they're not very credible, I feel like that kind of throws a little bit of logic out the window'' (P9), and
``I don't want to read people who are just saying [Design Proposals] with emotion, with no actual basis'' (P13).

In general, participants reflected the probe's functionality ``would just make it easier for me to find what I want to find'' (P2), especially if they were ``looking for specific types of information, like maybe within a topic'' (P5).
Some participants highlighted this focused search strength in context with their perception of large discussion forums as ``noise'' with ``long threads that are just pointless'' (P1), so that they don't have to ``manually go through all the comments, because there's a specific type of comment I'm looking for''
(P16) and ``saves you time from having to read everything'' (P5).

\subsubsection{Learning more about specific users or design perspectives via follow-up research both within and outside the discussion.}
While informative, some research from discussion forums wasn't inherently conclusive for participants, but the insights were helpful for guiding their future research.
In order to build from early-stage insights from discussion forums, participants recognized opportunities for more iterative research and design development via forums, in the form of interaction loops between researchers and commenters for more targeted, ongoing discussions.
P10 mentioned wanting more of a ``collaborative model'' to help researchers identify where to ask users for clarification and follow up to ``have that back and forth interaction'' in forums.
Other participants observed an opportunity in the probe's discussion thread for them to post more specific, direct prompts and even questionnaires for commenters to respond to, highlighting the potential for forums to foster targeted re-prompting or starting new discussions in an ongoing thread, even with complementary methods outside of traditional discussion forums.

In addition to recognizing follow-up interactions within the discussion forum, participants also recognized the power of understanding where to conduct follow-up analysis outside of the forum, mentioning ``it can help me make a better informed analysis of where I might be most concerned with'' (P21) and ``I can immediately, like, prioritize where I want to take my next step'' (P20), such as to ``look for different research papers that are highlighting these different topics [...] to give me a general idea of where to start to look to improve these systems'' (P7).

\subsection{RQ2: What barriers do researchers perceive around using LLMs for forum-based community research?}
Participants' understanding of online discussions was limited by the forum platform's perceived limitations, namely reconciling commenters' authenticity against their apparent expertise. 
In addition,
participants expressed limited trust in AI's ability to capture nuanced human insights, leading them to desire more structured control over LLM functionality for specific research goals.

\subsubsection{Uncertain user representativeness calls for more demographic context.}
Most participants acknowledged they would be wary of comments representing inaccurate information or opinions, without clear insight into commenters' backgrounds and intents.
For example, P7 mentioned ``the specific community that I'm looking at isn't necessarily a legitimized pool of people that I feel like might have the expertise to be talking about this.'' Other participants acknowledged forums may not be representative of wider target audiences.
Similarly, participants were wary of commenters' intentions on public forums, generally not trusting ``troll'' or ``cynical'' language in comments (P20), with some participants explicitly screening out comments ``phrased in almost a bad faith way'' (P20). While forums are accessible to many people and thus allow diverse inputs, it is difficult to control for the commenters' usage and intent.
Participants' wariness of commenters' intentions led them to desire more demographic information about commenters in order to contextualize their comments. As P21 put it, ``I would hope to be able to get some sense of who is responding with the comments [...] we definitely want to know who our population is [...]
I would want to see demographics.''

At the same time, however, participants also mentioned how the privacy and accessibility of anonymous forums can be at odds with this demographic information.
For example, P6 mentioned, ``I want to know who probably leaves these comments. For example, are they like experts [...] maybe it's hard to, like, infer from the anonymous forum,'' while P12 similarly stated, ``knowing how old they are, like maybe gender or something could be a good factor, but I don't even know if Reddit has that. I hope not, because I feel like that's an invasion of privacy.''
This commenter anonymity seemed tied in with the identified strengths of forums' low barrier of entry and freedom of expression, but participants desired at least some compromise on this anonymity in order to better understand their users and contextualize commenters' contributions.


\subsubsection{LLM summaries exclude details from the human voice.}
Participants appreciated how \frameworkshortname\ tagging made data easier to read, mentioning the data organization as a way to ``sort through information'' (P16), and required ``less cognitive load [...] because we're applying a structure [...]
I can get that information or digest it a lot better'' (P21).
Still, participants wanted to validate AI's tagging with source comments. One proposed way to do this was to show
``representative examples that can help me see what the comments for each [\frameworkshortname\ dimension]
look like'' (P6).

Since summarization necessarily excludes details, participants were even more wary of the lack of transparency and control of what details an LLM summarization excludes in the AI-Summarized Comment view of the probe.
In particular, participants noted skepticism about an AI being able to understand humans' emotions and lived experiences.
While the AI summary makes posts ``easier to parse'' (P3),
participants observed, ``I would honestly want to read the entire comment as a whole, like, at the end of the day, because I don't know if AI might have taken out information that I could have used''
(P16) or might ``lose some of the fine details that make the difference'' (P1),
``it's important to read what people are saying rather than just the AI [...] AI has its own understanding, but it's not living in the world, so it doesn't fully get it'' (P9),
``especially if we're dealing with something like emotions''
(P2).
A handful of other participants also mentioned the risks of AI ``hallucinations'', when the AI mentions ideas that aren't reflected in the source comments.
If LLM summarization was used, participants wanted to be able to validate the summary against the raw text, such as ``on top of this [raw] text'' (P6) or in ``its own bubble'' (P2), as long as it didn't ``replace the original text'' (P8).

Alternatively, participants recognized within-comment highlighting as a way to support their analyses without needing to rely on LLM summaries. They mentioned ``highlighting the relevant parts of these comments that pertain to this topic would kind of scaffold my search process'' (P1) as well as help ``extract quotes or that kind of wording [...] making it easier to get to the human parts of it'' (P8).

While specific perceptions of AI functionality varied, participants recognized that LLMs were not fully reliable because they couldn't control the AI's understanding and summarization of human anecdotes in forums, and it required their validation by reading the raw human-input comments.

\subsubsection{A need for more flexible levers to explore different analytical frames.}
While participants recognized value for targeted analysis,
they wanted more autonomy and control of analytical dimensions than current forum platforms and even the \frameworkshortname\ in our research probe allowed.
For example, several participants wanted to re-categorize or adjust the \topic\ breakdowns, such as by specifying ``early childhood education'' as a new category (P21). Similarly for \discourse, participants wanted the ability to add more specific tagging, such as ones designating if a discussion is ``on topic'' versus ``arguing on semantics'' (P8). A popular desire for \reasoning\ was more specific breakdowns of Ethos to filter on specific users' expertise, such as ``Hey, I'm a policymaker. Or hey, I'm a teacher, or a parent'' (P7).

Several participants mentioned opportunities to expand \frameworkshortname's overall tagging, potentially allowing both commenters and researchers to add custom tags to allow commenters to highlight their points they want to make and allow others to ``quickly find what [they] want to know, based on tags'' (P21).
This community-driven self-tagging suggestion could further contextualize the \frameworkshortname\ dimensions to avoid misrepresenting user populations, for instance as P21 noticed, ``I don't know if it makes sense for something else to be applying that [Reasoning] tag onto the response other than the user themselves [...] there may be cultural differences to how people express themselves.'' 

\begin{table*}[t]
    \centering
    \setlength{\leftmargini}{0.5em} 
    \begin{tabular}{@{}l p{4cm} p{4cm} p{4cm}@{}}
        \toprule
        & \textbf{\topic} & \textbf{\discourse} & \textbf{\reasoning} \\
        \midrule
        \multirow{3}{*}{\textbf{Breakdowns}} 
        & Comment content topics (e.g., School District Integration, Teacher Compensation)\vspace{0.5ex} 
        & Problem Statement, Design Proposal, Agreement, Disagreement\vspace{0.5ex}  
        & Logos (logic), Ethos (expertise/experience), Pathos (emotion)\vspace{0.5ex} \\
        \midrule
        \multirow{3}{*}{\textbf{Use Cases}} 
        & \begin{itemize}\setlength\itemsep{0pt}            
            \item Discover and search for specific subtopics (\textit{19})
            \item Identify common pain points and desired outcomes (\textit{16})
            \item Identify the breadth of topics for follow-up research (\textit{9})
        \end{itemize}
        & \begin{itemize}\setlength\itemsep{0pt}
            \item Inform targeted follow-up research, such as quote selection and argument fact-checking (\textit{15})
            \item Find popular Problems and Design Proposals (\textit{14})
            \item Find Disagreement signalling negative feedback for Design Proposals (\textit{12})
        \end{itemize}
        & \begin{itemize}\setlength\itemsep{0pt}
            \item Filter for valuable comments using a combination of reasonings with other dimensions (\textit{15})
            \item Ethos to find expert/niche experiences (\textit{14})
            \item Logos to find factual thinking and cited sources (\textit{14})
            \item Pathos to determine sentiment and bias (\textit{3})
        \end{itemize} \\
        \bottomrule
    \end{tabular}
    \caption{The breakdowns for each \frameworkname\ dimension, and the most commonly mentioned use cases for each dimension in our probe study. The number in parentheses denotes the distinct number of participants who mentioned each usage.}
    \label{tab:dim_goals}
\end{table*}

\section{DISCUSSION AND FUTURE WORK}
Our results highlight considerations and opportunities for injecting LLM functionality in forums and tools to support researchers' analytical needs.


\subsection{AI for Human Understanding, Beyond Summarization}\label{ssec:design_considerations}
Participants found \frameworkshortname\ breakdowns useful but were skeptical of the ability for AI summaries to capture human voice and context without easy verification.
Specifically, tagging supported faster validation against raw comments, while summaries required more effort to verify and prompted participants to read full comments anyway.
This implies that the current ability of LLMs to summarize human comments may not be sufficient for researcher understanding.

Future LLM-assisted tooling should integrate with grounded researcher processes and human agency \cite{jiang2021serendipity} rather than abstracting away the raw data with summaries, such as by highlighting relevant passages, surfacing representative quotes, and customizing transparent and evaluable LLM criteria. LLM functionalities should not replace researcher analysis, but they can help focus and validate the analysis.

\subsection{Tensions Between Open Communication and Analytical Utility}
Participants highlighted forums' low participation barriers and anonymity as strengths to encourage diverse collective discourse, but they wanted more context about who commenters were and whether inputs were representative or given in good-faith,
which are beyond typical forum platform affordances. Even if a commenter self-identifies their experience and \reasoning, their anonymity makes it difficult for researchers to confirm their standing in the community and the alignment of their underlying motives with analytical needs.

Future analytical forum platforms and tools should balance open expression with ways to contextualize expertise and intent, for example by surfacing contribution patterns or verified professions without exposing personal identities.
Based on our observation of participants' skepticism of LLM summaries, this kind of transparency will likely become more valuable as researchers use more LLM tools and communities become more aware of AI usage on forums \cite{lloyd2025aiuse}.
Previous work showed that anonymity makes people more willing to share their personal opinions \cite{wu2018anon}, but de-anonymizing social comments can also benefit community discourse \cite{omernick2013anon}, independent of downstream research value. In any case, communities have different values \cite{weld2024reddit} and warrant situated ethical approaches from researchers \cite{gliniecka2023ethics} to determine the right level of analytical transparency in a particular forum.


\subsection{Opportunities for Deeper Community Insights}
Forums enabled both analytical exploratory breadth and targeted depth.
Furthermore, participants particularly valued the potential for follow-up research, both within forums as conversational loops and beyond forums for informing complementary research.
Future tools could better support follow-up research, such as scaffolding researcher comments in discussions for deeper context, enabling functionality to triangulate forum insights with complementary primary research sources like interviews, and allowing commenters to voluntarily disclose contextual demographics or quickly add tags or reactions to specific researcher inquiries \cite{ko2022ugl}.

Naturally, forum analysis tools also need to account for potential observer effects and resource costs, as well as ethical concerns around public sharing spaces, confidentiality, and informed consent \cite{smedley2021practical}. Analyzing forums for research also warrants situated ethical considerations for users in these communities for potential social consequences \cite{gliniecka2023ethics}, especially with the growing popularity of AI research tools.

\section{LIMITATIONS}
Since we analyzed our probe on a single subreddit thread with 21 participants primarily from our local institution,
future work with broader samples and varied discussions would be crucial for understanding the generalizability of our findings.
Similarly, while the \frameworkname\ elicited grounded insight in our study, its validation as an analytical paradigm would require future work with broader samples.
Given the evolution of LLMs, we used a static prototype populated with human-labeled data to assure coherence but allowed participants to imagine the data resulted from  LLMs to elicit general perceptions rather than evaluate a specific model.
While our study did not systematically evaluate all modern LLM workflows or model consistency, we focused on the technical feasibility of one of these technologies (GPT-5-mini) to demonstrate its potential for analyzing discussion forums, and we frame our findings on human behaviors and results in the hopes of generalizing our insights to human interaction with future technologies regardless of the state of the technology.

\section{CONCLUSION}
We analyzed a real-world forum thread, synthesized an exploratory analytical framework from related work, validated preliminary LLM feasibility for extracting thread dimensions, and conducted design probe interviews with 21 researchers for their reflections on LLM-assisted forum analysis.
Forums can support a variety of research goals, but opportunities are moderated by a tension between anonymous participation and demographic understanding, LLM semantic limitations and researchers' trust in LLMs, and researchers' needs for flexible, source-grounded analysis sensemaking.
Our empirical results inform design implications for analyzing discussion forums to understand communities' collective intelligence.
\bibliographystyle{ACM-Reference-Format}
\bibliography{forum-bib}

\appendix
\section{PROBE STUDY INTERVIEW GUIDE}\label{append:interview_guide}
Semi-structured 1-hour online interview (see \autoref{sec:design_probe}). Interviewers asked follow-up questions beyond this outline.

\subsection{Warm-Up Questions (5 min)}
\begin{itemize}
    \item What online discussion forums have you used, and why?
    \item \textit{[If applicable]} What kinds of threads do you post, and how many comments do they get?
    \item \textit{[If applicable]} What qualitative data have you analyzed? What tools did you use?
\end{itemize}

\subsection{Baseline / Pre-Probe (15 min)}
Imagine you were a policymaker and posted this thread \textit{[share subreddit thread URL]} to help inform your design decision about this topic, and you got responses in the form of a subreddit thread of comments from your target users. What kinds of information would you hope to get from this data? Explain how you would get that information.

Feel free to take a couple minutes to look at the data to try to find that information you wanted. Please think aloud on what you do and what you observe as you do this.

\begin{itemize}
    \item Reflecting back on this, to what extent did you find the information you wanted to discover? Were there any challenges that made this difficult for you?
    \item Did you observe anything else with the data that informed your potential decision about this topic?
\end{itemize}

\subsection{Probe Tasks (30 min)}
\begin{itemize}
    \item Let’s say an AI was able to read all the discussion comments and came up with these topics, comprising the listed subtopics, that were mentioned by commenters \noindent\textit{[show Topic Selection]}. If you wanted to investigate information about these topics, you would be able to see data broken out by each high-level topic like this: \textit{[show Top-Level Topic Dashboard]}
    \item As a policymaker wanting to inform your design policy decision about the overall topic of K--12 education, how relevant or useful might this information be?
    \item Would you say these topics are organized in a sufficiently meaningful way? If not, how might you want to change or choose topics to look into?
\end{itemize}

\noindent\textit{[For each of the following features, provide examples and switch between granularity views:]}

\begin{itemize}
    \item Imagine you were able to see information about this discussion data’s \textit{[insert feature description]}. It might look something like this: \textit{[show Topic Overview of feature]}
    \item This is a top-level overview of the Topics, with a breakdown of the \textit{[discourse role / reasoning type]} for each topic.
    \item If you wanted to see particular comments…    \textit{[show Comment View feature screenshot]} \\
    …this is the low-level view which shows individual comments which you can read the raw text from and add filters on \textit{[topic / discourse role / reasoning type]} to view.

    \item As a policymaker wanting to inform your design decision about this topic, what kinds of information do you expect to get from these views?
    \item How relevant or helpful might this information be for your decision-making process? What might you hope to uncover with this feature?
    \item Is there any additional functionality you would want to see with this feature? How would that be helpful for you?
\end{itemize}

\subsection{Post-Probe Closing (10 min)}
\begin{itemize}
    \item What information would be most helpful for making your policy design decision?
    \item From the additional information we showed you, which aspects seemed most and least helpful for your decision-making?
    \item From the additional information we showed you, how easy or difficult might it be to help you find or uncover the information that would be most helpful for your policy design decision?
    \item Any other information that would be useful for your decision-making?
\end{itemize}

\section{TOPIC CODEBOOK}\label{append:topic_codebook}

Two researchers independently developed and refined this codebook through iterative discussion (see \autoref{ssec:reddit_dataset}).

\begin{table*}[htbp!]
\renewcommand{\arraystretch}{1.05}
\begin{tabular}{@{}p{5cm}p{9.5cm}@{}}
\toprule
\textbf{Category} & \textbf{Definition (Improving K--12 in U.S.)} \\
\midrule
School District Integration & Integrating schools across district or socioeconomic boundaries (e.g., busing, redistricting) as a way to equalize opportunity \\
Invest \$ in Students & Increased funding or resources directed to students (e.g., classroom materials, school facilities, extracurricular support) \\
Decentralized Education Systems & Shifting authority to local/state levels; fragmentation across states as shaping educational quality \\
Invest in Pre-K & Early childhood education (e.g., universal pre-K, expanding preschool access) as a foundation for later learning \\
Privatization & Private schools, vouchers, or charter expansion \\
Adjust School Resources (within and for) & Reallocating or restructuring resources within schools or districts (e.g., staffing, materials, standards) \\
Political Bias & Political influence or bias in curriculum, standards, or teaching \\
School Power / Teacher Agency & Giving teachers or schools more disciplinary authority. \\
Adjust School Scheduling / Timing & Adjusting calendars or daily schedules (e.g., start times, summer breaks) for learning and student well-being \\
Adjust Student Assessment Metrics & Reforming assessment systems (e.g., standardized testing, alternative metrics) \\
Professional Development for Teachers & Training, certification, or ongoing teacher development \\
Teacher Compensation & Raising or restructuring teacher pay to improve retention and teaching quality \\
Overhaul Curriculum & Revising, modernizing, or diversifying curriculum (e.g., STEM, vocational tracks, charter alternatives) \\
Promote Parental Involvement (Homeschooling) & Stronger parental engagement in education, including homeschooling. \\
\bottomrule
\end{tabular}
\caption{Topic categories from our manual codebook of a subreddit thread on education policy, with their defined inclusion criteria}
\end{table*}

\end{document}


\newcommand{\frameworkname}{\texttt{Topic}-\texttt{Discourse}-\texttt{Reasoning Framework}}
\newcommand{\frameworkshortname}{\texttt{TDR Framework}}
\newcommand{\edit}[1]{\textcolor{blue}{#1}}
\newcommand{\sssec}[1]{\textbf{#1}}
\newcommand{\tabitem}{~~\llap{\textbullet}~~}
\newcommand{\topic}{Topic}
\newcommand{\topics}{Topics}
\newcommand{\discourse}{Discourse Role}
\newcommand{\discourses}{Discourse Roles}
\newcommand{\reasoning}{Reasoning Type}
\newcommand{\reasonings}{Reasoning Types}

\maketitle

\section{LLM PROMPT TEMPLATES}\label{append:llm_prompts}
This section provides the custom prompt templates we applied for our workflows. Standard prompts adapted directly from \textsc{LLooM} \cite{lam2024lloom} (e.g., for bullet-point distillation and graded scoring) are not repeated here.

\subsection{Category Definitions for LLM Scoring}
The following definitions were applied within the \textsc{LLooM} categorization prompt format (\ref{ssec:scoring_filtering}). Each category was defined with inclusion criteria and representative examples to guide the model:

\begin{itemize}
    \item \textbf{Problem Statement}: A comment describes an issue, challenge, or shortcoming related to the subreddit question. The issue must be a real concern rather than a hypothetical problem, but a comment may still qualify even if it also suggests a solution.
    \item \textbf{Design Proposal}: A comment proposes a solution to the issue or question being discussed. Even brief suggestions count as design proposals.
    \item \textbf{Logos Reasoning}: A comment uses logical reasoning, such as facts, causal links, or structured arguments, to support its point.
    \item \textbf{Ethos Reasoning}: A comment references personal or professional experience, or the perspective of an authority, to strengthen its argument (including anecdotes or expertise).
    \item \textbf{Pathos Reasoning}: A comment uses emotional appeals—such as sympathy, outrage, or inspiration—to persuade the reader.
    \item \textbf{Disagreement}: A comment explicitly expresses disagreement with the parent comment it replies to.
    \item \textbf{Agreement}: A comment explicitly expresses agreement with the parent comment it replies to.
\end{itemize}

\subsection{Bottom-Up Workflow Prompts}

\subsubsection{High Specificity}
\begin{Verbatim}[breaklines=true]
review_merge_prompt = """
I have this set of themes generated from text examples:
{concepts}

These concepts are generated from these bullet points around how to improve k-12 US education system:
{bullets}. 

Please identify any PAIRS of themes that are similar or overlapping that should be MERGED together. 
Please respond ONLY with a valid JSON in the following format with the original themes and a new name and prompt for the merged theme. Do NOT simply combine the prior theme names or prompts, but come up with a new 2-3 word name and 1-sentence ChatGPT prompt. If there no similar themes, please leave the list empty;

{
    "merge": [ 
        {
            "original_themes": ["<THEME_NAME_A>", "<THEME_NAME_B>"],
            "merged_theme_name": "<THEME_NAME_AB>",
            "merged_theme_prompt": "<THEME_PROMPT_AB>"
        },
        {
            "original_themes": ["<THEME_NAME_C>", "<THEME_NAME_D>"],
            "merged_theme_name": "<THEME_NAME_CD>",
            "merged_theme_prompt": "<THEME_PROMPT_CD>"
        }
    ]
}
"""
\end{Verbatim}

\subsubsection{Mid Specificity}
\begin{Verbatim}[breaklines=true]
review_merge_mid_prompt = """
I have this set of themes generated from text examples:
{concepts}

These concepts are generated from these bullet points around how to improve k-12 US education system:
{bullets}. 
    
Please identify pairs of themes that share a moderate overlap and should be merged into mid‐level specificity groups. 
Merge only when two themes have clear commonalities but keep others intact. 
Do NOT simply combine the prior theme names or prompts, but come up with a new 2-3 word name and 1-sentence ChatGPT prompt. 
Please respond ONLY with a valid JSON in the following format with the original themes and a new name and prompt for the merged theme.
If there no similar themes, please leave the list empty:
{
    "merge": [
        {
            "original_themes": ["<A>", "<B>"],
            "merged_theme_name": "<AB>",
            "merged_theme_prompt": "<Prompt for AB>"
        }
    ]
}
"""
\end{Verbatim}

\subsubsection{Low Specificity}
\begin{Verbatim}[breaklines=true]
review_merge_low_prompt = """
I have this set of themes generated from text examples:
{concepts}

These concepts are generated from these bullet points around how to improve k-12 US education system:
{bullets}. 

Please identify pairs of themes that should be broadly merged into high‐level, overarching categories. 
Do NOT simply combine the prior theme names or prompts, but come up with a new 2-3 word name and 1-sentence ChatGPT prompt. 
Please respond ONLY with a valid JSON in the following format with the original themes and a new name and prompt for the merged theme.
If there no similar themes, please leave the list empty:
{
    "merge": [
        {
            "original_themes": ["<C>", "<D>"],
            "merged_theme_name": "<CD>",
            "merged_theme_prompt": "<Prompt for CD>"
        }
    ]
}
Do NOT include any trailing commas after the last field or array element.
"""
\end{Verbatim}

\subsection{Top-Down Workflow Prompts}

\subsubsection{Low Specificity}
\begin{Verbatim}[breaklines=true]
topdown_low_prompt = """
You are given a list of bullet points extracted from a discussion about how to improve k-12 US education system

Your task is to identify **around {expected_count} broad and low-specificity** themes that summarize the key ideas.

### Specificity Level Definitions:
- Low specificity: Broad, general categories that group many ideas.
- Mid specificity: Moderately detailed groupings that capture meaningful sub-themes.
- High specificity: Narrow, precise topics describing specific issues or proposals.

Respond ONLY with a valid JSON object under the key "concepts". No extra text, no trailing commas:

{
  "concepts": [
    {
      "name": "<THEME_NAME>",
      "prompt": "<THEME_PROMPT>"
    }
  ]
}

### Bullet Points:
{bullets}
"""
\end{Verbatim}

\subsubsection{Mid Specificity}
\begin{Verbatim}[breaklines=true]
review_unmerge_mid_prompt = """
I have these low-specificity themes:
{concepts}

And here are all the bullet points:
{bullets}

Please split them into exactly {expected_count} mid-specificity subthemes.  
- Decide which parent themes to break apart.  
- For each parent you split, output its “parent_theme” plus a “subthemes” list.  
- Each subtheme needs a concise 2–3 word “name” and a one-sentence “prompt.”  
- If a parent theme does not need splitting, omit it.  
- If no themes need splitting, return "unmerge": [].

Respond ONLY with a single JSON object under the key "unmerge":

{
  "unmerge": [
    {
      "parent_theme": "<LOW_THEME>",
      "subthemes": [
        { "name": "<MID_NAME_1>", "prompt": "<MID_PROMPT_1>" },
        { "name": "<MID_NAME_2>", "prompt": "<MID_PROMPT_2>" }
      ]
    }
  ]
}
"""
\end{Verbatim}

\subsubsection{High Specificity}
\begin{Verbatim}[breaklines=true]
review_unmerge_high_prompt = """
I have these mid-specificity themes:
{concepts}

And here are all the bullet points:
{bullets}

Please split them into exactly {expected_count} high-specificity subthemes.  
- Decide which parent themes to break apart further.  
- For each parent you split, output its “parent_theme” plus a “subthemes” list.  
- Each subtheme needs a concise 2–3 word “name” and a one-sentence “prompt.”  
- If a parent theme does not need further splitting, omit it.  
- If no themes need splitting, return "unmerge": [].

Respond ONLY with a single JSON object under the key "unmerge":

{
  "unmerge": [
    {
      "parent_theme": "<MID_THEME>",
      "subthemes": [
        { "name": "<HIGH_NAME_1>", "prompt": "<HIGH_PROMPT_1>" },
        { "name": "<HIGH_NAME_2>", "prompt": "<HIGH_PROMPT_2>" }
      ]
    }
  ]
}
"""
\end{Verbatim}

\section{LLM Workflow Implementation and Evaluation}

To evaluate the preliminary feasibility of our LLM workflow with \frameworkname, we utilized the same subreddit thread used for the research probe \cite{redditthread:online} to observe how reliably the workflow generated \topics\ from the discussion and evaluate how well the workflow could classify comments under each dimension of \frameworkshortname\ compared to existing AI classification methods. This section outlines our implementation logic and evaluation details; please reach out directly to the first two authors if you want to request access to the raw code scripts.

\subsection{Topic Generation}
The first technical challenge was to inductively generate \topics\ from a discussion, without a predetermined codebook.

\subsubsection{Implementation Details}\label{app:topic_gen_implementation}
We explored two ways to derive clustered topics: a bottom-up workflow and a top-down workflow.
Both approaches begin by applying an LLM to distill raw comments into concise bullet points. This step helps reduce redundancy, highlights ideas most relevant to the research question, and produces a tractable representation of the dataset for downstream analysis.
We adopted this strategy based on prior work that demonstrates the effectiveness of LLM-based summarization for converting conversational data into analyzable units \cite{turbeville2024llm, lubos2025towards, asthana2025summaries, viswanathan2024large}.   

\sssec{Top-Down Workflow.}\label{ssec:topic_top_down}
The top-down workflow starts from a single seed topic (the overall theme) and progressively unmerges it into more specific topics. At the high-level (lowest specificity) stage, the LLM receives (i) the seed topic, (ii) the full bullet set, and (iii) a manually specified expected topic count \(N_{\text{high}}\); it is instructed to produce approximately \(N_{\text{high}}\) broad topics, each with a name and a one-sentence description. At the successive mid-level stage, the LLM receives (i) the seed topic, (ii) the high-level topics (names and descriptions) as candidate parents to split, (iii) the full bullet set, and (iv) the expected count \(N_{\text{mid}}\); it selects which parents to split and returns a more specific set targeting \(N_{\text{mid}}\). At the final, low-level (high specificity) stage, the same procedure is repeated using the mid-level topics as candidate parents and the expected count \(N_{\text{low}}\). Expected counts \(N_{\text{high}}, N_{\text{mid}}, N_{\text{low}}\) can be either set by the human researchers or derived from comment numbers; in this study, defaults of approximately \(10\%\), \(5\%\), and \(2.5\%\) of the total number of comments are used for the high-, mid-, and low-levels, respectively. 

\sssec{Bottom-Up Workflow.}\label{ssec:topic_bottom_up}
The bottom-up workflow instead begins with the distilled bullet points. First, bullet points are embedded and clustered to form semantically coherent groups; next, the LLM is prompted to name each cluster and provide a one-sentence description, yielding an initial set of  candidate topics. Subsequently, to match the expected topic counts for the low, mid, and high-level topics (defined as in \autoref{ssec:topic_top_down}), candidates are refined via LLM-guided merging. Merging proceeds iteratively: in each round, the model receives the current topic list (name and prompt) together with the full bullet set and proposes pairwise merges of similar topics, each with a merged name and prompt. Inspired by \textsc{LLooM} \cite{lam2024lloom}, any given topic may participate in at most one merge per round to avoid conflicting or chained merges. Rounds continue until the number of topics at the current level is at or below the expected count for that level, or until no further merges are proposed. Finally, the resulting set is carried forward to the next, broader level.


\subsubsection{Evaluation Methods}\label{app:topic_gen_eval}
We evaluated two structured variants of our approach, \textit{top-down} (\autoref{ssec:topic_top_down}) and \textit{bottom-up} (\autoref{ssec:topic_bottom_up}), rather than an LLM with single direct prompting baseline. We chose this method since previous work has established that scaffolded multi-step prompting with intermediate artifacts outperforms direct prompting on complex tasks and provides greater transparency and control \cite{wei2022chain, grunde2025designing, chen2023program}. Our evaluation focused on if and how \topic generation may vary by top-down or bottom-up workflows while holding other LLM configurations constant.

First, we evaluated each workflow's concept coverage of the ground truth \topics we manually labeled, similar to previous work \cite{lam2024lloom, rao2024quallm}. Looking at each workflow's low-level (highest-specificity) concepts, two authors independently semantically matched each concept with one of the manual topics and then calculated the percentage of manual topics that had a match with an AI-generated concept. If a low-level AI concept did not relate to any manual topic, it was discarded; if it covered multiple different manual topics,
it was assigned to the topic it most associated with.
The coverage percentages were averaged between the two labelers' evaluations.
After checking coverage, one author also conducted inductive open coding on the topics generated, observing differences and similarities between the two AI workflows' results. 

\subsubsection{Evaluation Results}\label{app:topic_gen_results}

Our top-down LLM workflow's low-level concepts mapped to an average of 82\% of our manually-coded topics in the discussion thread, with both annotators agreeing it failed to address the topics of `Adjust School Scheduling and Timing' and `Promote Parental Involvement'. On the other hand, our bottom-up approach mapped an average of 89\% of the manual topics, with both annotators agreeing it missed the `Invest in Pre-K' topic. This means the bottom-up workflow appeared to capture slightly more of our ground truth topic labels from the dataset.

The bottom-up and top-down workflows both generated detailed topics and descriptions, each also having the most concepts map to the manual label of `School District Integration', which is reasonable since that topic had the most mentions in our dataset. Both workflows also had similar levels of specificity in their concepts, each with three concepts lacking unambiguous details, such as the generated concept of `Intervention Systems: Pair accountability with transparent interventions, supports, and consequences that help struggling schools improve rather than simply punish them'.
However, looking at the mapped concepts and their AI-generated descriptions, the top-down concepts had more homogeneity of the concepts within their mapped manual labels, with four pairs of nearly identical concepts versus the bottom-up's two pairs of closely similar concepts. For example, the top-down workflow's low-level concepts of `Busing \& Redistricting' and 
`Access \& Transport' both detail targeted transportation options to reduce inequity, within the broad human-labeled category of `School District Integration'.

Both workflows yielded similarly specific concepts, but the bottom-up workflow seemed to detail slightly less ambiguity in its concept descriptions and cover more variability within higher-level topics, in addition to covering more of the manually-labeled topics.

 \subsection{\frameworkshortname\ Categorization Workflow} \label{sec:categorization_workflow}
 The second technical challenge was to accurately label each comment in a discussion within each dimension of the \frameworkname.

\subsubsection{Implementation Details}\label{app:dim_implementation}
To automate the categorization of discussion comments across all three dimensions of \topic, \discourse, and \reasoning, we employed an LLM-based categorization system that classifies textual data into predefined categories, which consisted of two main phases: (1) category specification (with prompt construction), and (2) LLM scoring with threshold-based filtering.

\sssec{Category specification.} \label{ssec:cat_spec}
For comparative evaluation, we specified our dimension categories to be the manually-labeled dataset from our probe study. We also utilized the manual topic labels for our workflow, though LLM-generated topics could also be used.
For the \topic\ dimension, we also included inclusion criteria from our open coding of the topic labels as context for the LLM.
For each category across all three dimensions, we handpicked representative example comments. These definitions, criteria, and examples formed the category information used in subsequent steps in our LLM workflow.

\sssec{LLM scoring and filtering.} \label{ssec:scoring_filtering}
This step was executed independently for each dimension. For each target comment, the LLM received (i) the category information (label names, inclusion criteria, and handpicked examples), (ii) the target comment text, and (iii) if the target is a reply, its parent comment (one level up) for conversational context. For \discourses, Agreement/Disagreement was queried only when a parent existed. Then, to represent nuanced degrees of fit between comments and categories across the three dimensions \cite{zhang2017discourse, kulkarni2022ctm}, we adopted the graded scoring pattern used in \textsc{LLooM} \cite{lam2024lloom}. Specifically, based on all inputs, the LLM was instructed to evaluate the target comment against all categories in the current dimension and return, for each category, a graded relevance in $\{A,B,C,D,E\}$ (where $A$ = strongly relevant and $E$ = not relevant) with a one-sentence rationale anchored to the inclusion criteria. Finally, human researchers set a global acceptance threshold $T$ after a sample review of the scoring results; for each dimension, we retain all category–comment pairs with relevance grade $\geq T$ and discard the rest.

\sssec{Technical Details}
We implemented the workflow in Python. For all steps mediated by LLMs in topic generation and categorization (\autoref{ssec:scoring_filtering}), we used \texttt{GPT-5-mini} via the OpenAI API with \texttt{ reasoning\_effort = high}. In the bottom-up topic generation path (\autoref{ssec:topic_bottom_up}), we encoded comments with \texttt{text-embedding-3-large} and clustered embeddings using \textsc{HDBSCAN}. For graded categorization (five-grade scheme $\{A,B,C,D,E\}$ as defined in \autoref{ssec:scoring_filtering}), we drew a 10\% random sample, compared model grades against manual codes in all three dimensions, and set an acceptance threshold \(T=\textbf{A}\). Prompts for bullet point distillation in topic generation and for LLM scoring in categorization were adapted from \textsc{LLooM} \cite{lam2024lloom}. All other prompt templates are provided in \autoref{append:llm_prompts}.

\subsubsection{Classification Evaluation}\label{app:dim_eval}
We framed the classifications of \discourse\
and \reasoning\
as multi-label classification problems. Using our manual topic labels as pre-defined categories, we also framed \topic\ classification as a multi-label classification problem. While evaluating topic classification on our manual labels doesn't necessarily capture the end-to-end behavior of our LLM process, we chose this framing for more control over evaluation and comparability with other models.

For classification models of comparison, we trained a simple logistic regression model based on TF-IDF vectorization of cleaned comment text, a BERT transformer model, and a RoBERTa transformer model.
We trained these models on 80\% of the manually labeled comments, evaluating with 5-fold stratified cross-validation and averaging results. We evaluated these models this way due to the relatively small dataset of a single discussion thread and to be more comparable with our one-shot LLM model, assuming the designer could manually label some comments to fine-tune the model. Similar to our LLM process, we fine-tuned the prediction thresholds for the BERT and RoBERTa models to maximize possible F1 scores.
We recognize it could be unreasonable for a designer to label 80\% of a thread's comments and fine-tune parameters for each discussion, but we made some effort to maximize the performance of these models to present a more conservative comparison.

We calculated our LLM workflow's and the baseline models' classification accuracies
against our manual annotations for each categorization of \topic, \discourse, and \reasoning, in a one-versus-all schema, similar to how previous work evaluated classification against manual topics \cite{rao2024quallm}.
We filtered out our secondary labels of Humor, Off-Topic, and Asking for Clarification from our \discourse\ analysis due to low relevance to our study goals around design insights and low presence in our manually-labeled dataset. 
Our manual dataset reflected imbalance in labels across all \frameworkshortname\ dimensions, for instance skewing more towards Disagreement in \discourse\ and Logical in \reasoning, so we calculated the macro F1 scores (the average of each label's precision and recall metrics) for each model to capture classification performance on minority classes.
For evaluative comparison, we also averaged our LLM workflow's F1 scores across the same 5-fold test splits used in the baseline models.

\subsubsection{Classification Evaluation Results}\label{app:dim_results}
Our LLM workflow consistently outperformed the comparison models for classifying all dimensions of \topic, \discourse, and \reasoning, addressing RQ2 by showing potential feasibility for LLMs to scale analyses with the \frameworkshortname\ at least in comparison with popular natural language processing models. The LLM workflow's \discourse\ classification had the highest macro F1 score of 0.48, and the \topic\ classification showed the greatest increase in F1 scores, with a macro F1 of 0.42 over the best baseline's score of 0.15, a 180\% relative increase. Macro F1 scores for all dimensions are shown in \autoref{t:accuracy_overview}.

\begin{table}
\centering
\begin{tabular}{c | c c c}
 \hline
  Model & Topic & Discourse Role & Reasoning Type \\
 \hline
 Our LLM Workflow & \textbf{0.42} & \textbf{0.48} & \textbf{0.42} \\ 
 RoBERTa & 0.15 & 0.31 & 0.29 \\
 BERT & 0.15 & 0.29 & 0.24 \\
 LogReg & 0.01 & 0.12 & 0.22 \\
 \hline 
\end{tabular}
\caption{Macro-averaged F1 scores for multilabel classification with our LLM workflow based on LLooM compared with logistic regression, BERT, and RoBERTa trained on 80\% of the data. Our LLM outperformed the others across all \frameworkshortname\ dimensions.}
\label{t:accuracy_overview}
\end{table}


\bibliographystyle{ACM-Reference-Format}
\bibliography{forum-bib}